\documentstyle[epsfig]{aipproc}
\def\NII{\hbox{N~$\scriptstyle\rm II\ $}}
\def\HI{\hbox{H~$\scriptstyle\rm I\ $}}
\def\msun{{\rm\,M_\odot}}
\def\msunits{{\rm\,M_\odot\,yr^{-1}\,Mpc^{-3}}}
\def\munits{{\rm\,M_\odot\,Mpc^{-3}}}
\def\ub{U_{300}-B_{450}}
\def\bv{B_{450}-V_{606}}
\def\vi{V_{606}-I_{814}}
\def\bi{B_{450}-I_{814}}

\begin{document}
\title{Cosmic Star Formation History}


\author{Piero Madau}
\address{Space Telescope Science Institute, 3700 San Martin Drive, 
Baltimore, MD 21218}

\maketitle

\begin{abstract}
I review some recent progress made in our understanding of galaxy evolution and
the cosmic history of star formation. Like bookends, the results obtained from
deep ground-based spectroscopy and from the Hubble Deep Field imaging survey
put brackets around the intermediate redshift interval, $1<z<2$, where
starbirth probably peaked at a rate 10 times higher than today. The steady
decline observed since $z\sim 1$ is largely associated with late-type galaxies.
At $z\gtrsim 2.5$, the Lyman-break selected objects may represent the
precursors of present-day spheroids, but appear, on average, quite
underluminous relative to the expectations of the standard early-and-rapidly
forming picture for spheroidal systems. The observed ultraviolet light density
accounts for the bulk of the metals seen today in ``normal'' massive galaxies. 

\end{abstract} 

\section*{introduction}
The knowledge of the star formation rate (SFR) throughout the universe as a
function of space and time is one of the primary goal of galaxy formation and 
evolution studies. Key questions to be answered are as
follows: How can we identify high-$z$ galaxies in deep CCD surveys? Are they
forming stars more rapidly than quiescent spirals at $z\sim 0$? Are
they obscured by dust in analogy with luminous IRAS starbursts? Is
there a characteristic epoch of star and element formation in galaxies? How
does the distribution of SFR evolve with redshift? Do spheroids form early and
rapidly? What is the origin of disk galaxies?

Two complementary approaches can be used to shed light on these questions. One
is to study the resolved stellar populations of the Milky Way and nearby 
galaxies and infer their evolutionary history from fossil records -- examples 
of this approach are nuclear cosmochronology, the color-magnitude diagram of
globular clusters, the cooling sequence of white dwarfs. The other is to
systematically observe galaxies at increasing cosmological lookback times, and
reconstruct the history of stellar birthrate {\it directly}. In this talk I
will review the broad picture that has recently emerged from the
``direct'' method, focusing on what can be learned 
from integrated quantities over the entire population, rather than from a
detailed study of morphologically-distinct samples whose physical significance
remains unclear.  I will show how the combination of HST deep imaging and
ground-based spectroscopy offers now an exciting first glimpse to the history
of the conversion of neutral gas into stars in field galaxies. 

In the following, I will make use of the fact that the 
UV-continuum emission from a galaxy with significant ongoing star formation
is dominated by short-lived massive stars, and is therefore nearly
independent of the galaxy history. In all the transformation from UV luminosity
to SFR, a Salpeter IMF including stars in the $0.1<M<125\msun$ range with solar
metallicity will be assumed. For a Scalo IMF -- less rich in massive stars --
in the same mass range the conversion factor is about 3 times larger. All
magnitudes will be given in the AB system, and a flat cosmology
with $q_0=0.5$ and $H_0=50\,$km s$^{-1}$ Mpc$^{-1}$ will be adopted.

\section*{the local universe}
The Universitad Complutense de Madrid (UCM) objective-prism survey for
H$\alpha$-emitting objects [1] provides an ideal tool for studying
the properties of star-forming galaxies at low redshift. The sample consists of
about 250 sources in 500 deg$^2$ ($z<0.045$) with EW(H$\alpha+[\NII])>10$\AA,
and is dominated by intermediate- to low-luminosity late-type galaxies. The
line emission comes largely from the nuclear regions, and has been corrected
for reddening using the Balmer decrement. We can then use case B recombination
theory to relate the H$\alpha$ line luminosity to the rate of production of
ionizing photons, and population synthesis models [2] to estimate the
instantaneous SFR,  $L({\rm H}\alpha)=3\times 10^{41}\times$ SFR ergs s$^{-1}$,
where SFR is in units of $\msun\,$yr$^{-1}$. A Schechter function with
$\alpha=-1.3\pm 0.2$, SFR$^*=4.7\pm 0.8\msun\,$yr$^{-1}$, and
$\phi^*=10^{-3.2\pm 0.2}\,$Mpc$^{-3}$, shown in Figure 1, provides a good fit
to the present-epoch ``SFR function'' -- which describes the number of
star-forming galaxies as a function of their ongoing SFR -- in the range
between 0.1 and $10\msun\,$yr$^{-1}.$ Integrating over all luminosities, 
the total SFR density is
\[
{\dot \rho_*}=10^{-2.4\pm 0.2}\, {\rm M_\odot\,yr^{-1}\,Mpc^{-3}}.
\]
A similar value can be derived starting from the observed luminosity function
(LF) in the $B$-band [3], applying a luminosity-weighted average
color of $\langle 2800-4400\rangle_{\rm AB}=2.05\,$mag [4] to the
$B$-band luminosity density, and then converting the UV flux into an
instantaneous SFR. 

\begin{figure}[b!] 
\centerline{\epsfig{file=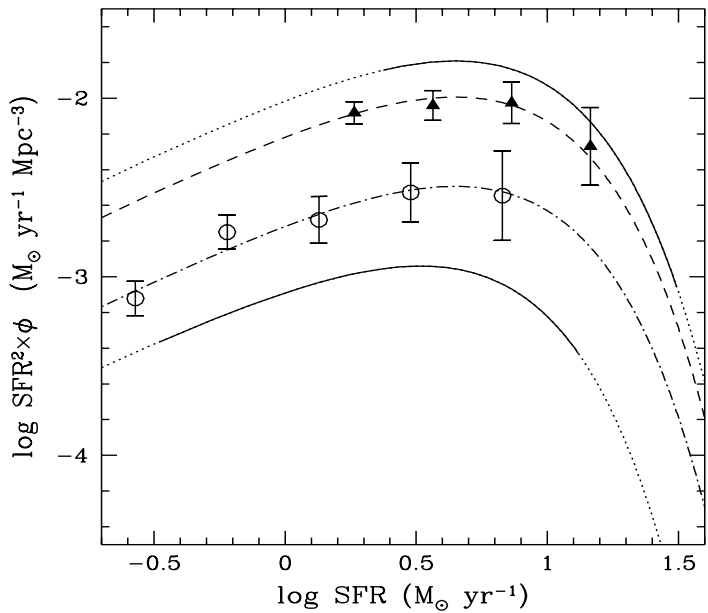,height=4.0in,width=4.5in}}
\vspace{10pt}
\caption{Distribution of star formation rates at different redshifts. The solid
line represents the Schechter function fitted to the data points, and the
dotted line its extrapolation. {\it Top curve}: SFR density at $0.75<z<1$. 
{\it Bottom curve}: SFR density today. The dashed line depicts the same
$0.75<z<1$ Schechter function with $\phi^*$ lower by a factor of 1.6. This
provides a good fit to the HDF points ({\it filled triangles}) at
$\langle z\rangle=2.75$. {\it Dot-dashed line:} same for blue dropouts.
} 
\label{fig1} 
\end{figure}

\section*{surveys to $z\sim 1$}
The recent completion of several comprehensive ground-based redshift surveys
[5--7] has significantly improved our undestanding of
the evolution of field galaxies to $z\sim 1$. A similar trend is  seen by the
various groups, namely the rapid evolution -- largely driven by late-type
galaxies -- of the LF with lookback time. In particular, from the\ marked
increase with redshift of the comoving luminosity density at 2800\AA\, observed
in the Canada-France Redshift Survey (CFRS) of 730 I-selected galaxies with
$17.5<I_{\rm AB}<22.5$ [5], the total (integrated over all
luminosities) rate of star formation per unit volume is
\[
{\dot \rho_*}=10^{-1.3\pm 0.15}\left({1+z\over 1.875}\right)^{3.9\pm 0.75} 
\, {\rm M_\odot\,yr^{-1}\,Mpc^{-3}}
\]
over the range $0<z<1$. The distribution of SFR in the interval $0.75<z<1$ is
shown in Figure 1. It has been derived from the rest-frame $B$-band LF at that
epoch [5], by applying a luminosity-weighted average color of $\langle
2800-4400 \rangle_{\rm AB}=1.3\,$ mag [4]. A Schechter function with
$\alpha=-1.28$, SFR$^*=6.2\,$M$_\odot\,$yr$^{-1}$, and $\phi^*=10^{-2.17}\,
$Mpc$^{-3}$ provides a good fit to the SFR function in the
$2.5-30\,$M$_\odot\,$yr$^{-1}$ range. 

A comparison with the present-day distribution shows the sign of a strong
density evolution -- a rapid increase in $\phi^*$. If fitted by an exponential,
the ten-fold increase in the volume-averaged SFR over the past 8 Gyr implies an
e-folding time of $\tau_{\rm SFR}=3.2^{+0.8}_{-0.5}$ Gyr.  The evolution is
strongly differential with color, with the LF of galaxies redder than a
present-day Sbc showing very little change with cosmic time. A morphological
analysis shows that one can identify disk-dominated galaxies (with bulge/total
luminosity $<0.5$) with the blue population whose LF is evolving rapidly, and
that galaxy disks at $\langle z\rangle=0.7$ have mean rest-frame central
surface brightness $\sim 1.6$ mag brighter than their local counterpart [8]. 

\section*{lyman-break galaxies}
Recently, the study of field galaxies has been pushed to much earlier lookback
times by the development of a broad-band color technique for identifying
galaxies at $z\sim 3$ [9,10]. At this redshift, the
Lyman-continuum break at 912\AA\ -- which arises from a combination of the
intrinsic discontinuity in the spectra of hot stars, the opacity of the galaxy
to its own ionizing radiation, and the ubiquitous effect of \HI absorption in
the intergalactic medium along the line of sight [11] -- passes
through the $U$-band, resulting in an unmistakable color signature for distant
star-forming galaxies: a very red $U-B$ color, combined with colors in
longer-wavelength filters that are much bluer.  Recent deep spectroscopy with
the W. M. Keck telescope has shown the high efficiency of such color selection.
Steidel and collaborators [12,13] have confirmed the
redshifts of $\sim 100$ galaxies at $2.9<z<3.4$ in four high-latitude fields,
and established that the technique is $>90\%$ reliable in going from
color-selected ``UV dropouts'' to confirmed high-$z$ galaxies, the only
contaminants being halo subdwarfs.

The UV-continuum fluxes of Lyman-break galaxies imply star formation rates in
the range 4--$100\,$M$_\odot\,$ yr$^{-1}$, and a SFR density of 
\[
{\dot \rho_*}(4-100)=10^{-2.13}\, {\rm M_\odot\,yr^{-1}\,Mpc^{-3}}
\]
at $\langle z\rangle=3.25$. This is $6$ times higher than the rate of starbirth
today, but $3$ times lower than the corresponding value at $z\sim 1$ 
{\it over the same range of luminosities}, and show the existence of a peak at
intermediate redshifts in the volume-averaged SFR in bright galaxies. 

Few pieces of evidence may support the interpretation that such galaxies are
bulges and spheroids seen in early formation: a) their volume density and
rest-frame optical luminosities derived from $K$-band photometry are comparable
to those of bright galaxies today; and b) {\it HST}-WFPC2 images show the
presence of compact, relaxed ``cores'' which are few kpc in size, in analogy
with the bulges and cores of luminous galaxies at the present epoch [14].

\section*{the hubble deep field}
The Hubble Deep Field (HDF) imaging survey has been specifically designed with
the application and generalization of the UV dropout technique in mind. With
its unprecedented depth, reaching 5-$\sigma$ limiting AB magnitudes of roughly
27.7, 28.6, 29.0, and 28.4 in the $U_{300}$, $B_{450}$, $V_{606}$, and
$I_{814}$ bandpasses [15], and four-filter strategy in order to
detect Lyman-break galaxies at various redshifts, the HDF is now a key
testing ground for models of galaxy evolution. 

\bigskip {\bf Galaxy Counts} \qquad There are about 3,000 galaxies in the HDF,
corresponding to $2\times 10^6$ deg$^{-2}$ to $V\sim 29$. In all four
bands, the slope $\alpha$ of the differential galaxy counts, $\log N(m)=\alpha
m$, flattens at faint magnitudes, e.g., from $\alpha=0.45$ in the interval
$21<B<25$ to $\alpha=0.17$ for $25<B<29$. This feature cannot be due to the
reddening of distant sources as their Lyman break gets redshifted into the blue
passband, since the fraction of Lyman-break galaxies at $B\sim 25$ is only of
order 10\% (cf [16]). Moreover, an absorption-induced loss of sources
could not explain the similar flattening of the number-magnitude relation
observed in the $V$ and $I$ bands [17]. Rather, the change of slope
suggests a decline in the surface density of luminous galaxies beyond $z\sim
1$. 

Since, for $\alpha<0.4$, the extragalactic background light (EBL) is
dominated by object at the bright end of the luminosity range, the flattening
of the number counts has the interesting consequences that the galaxies that
produce $\sim 60\%$ of the blue EBL have $B<24.5$. They are thus bright enough
to be identified in spectroscopic surveys, and are known to have median
redshift $\langle z\rangle=0.6$ [5]. The quite general conclusion is
that there is no evidence in the number-magnitude relation for a large amount
of star formation at high redshift. Note that these considerations do not
constrain the {\it rate} of starbirth at early epochs, only the total
(integrated over cosmic time) amount of stars -- hence background light --
being produced. The most direct way to track the evolution of the SFR density
at early epochs is through a census of the HDF dropouts. 

\bigskip {\bf Ultraviolet Dropouts} \qquad New photometric criteria for robustly
selecting Lyman-break galaxies have been developed based on the HDF color
system. By simulating colors for an extremely wide range of model galaxy
spectra, the criteria have been tuned up to provide what appear to be largely
uncontaminated samples of star-forming galaxies at high redshifts
[18]. I have further refined them after the many redshift
measurements with Keck.

The $U_{300}$ passband -- which is bluer than the standard ground-based $U$
filter -- permits the identification of star-forming galaxies in the 
interval $2<z<3.5$. Galaxies in this redshift range predominantly occupy the
top left portion of the $\ub$ vs. $\bi$ color-color diagram because of the
attenuation by the intergalactic medium and intrinsic extinction. Galaxies at
lower redshift can have similar $U-B$ colors, but they are typically either old
or dusty, and are therefore red in $B-I$ as well. About $100$ ultraviolet
dropouts can be identified in the HDF which are brighter than $B=27$, about
25\% of the total. Of these, 17 have spectroscopically
confirmed redshift in the range $2.2<z<3.4$ [19,20]. Note
that, out of the $\sim 60$ galaxies in the HDF with known redshifts $z<2$
[21], no low-redshift interlopers have been found among the
high-redshift sample. The color-selection is illustrated in Figure 2. The $UBI$
criteria isolate objects that have relatively blue colors in the optical, but a
sharp drop into the UV. A ``plume'' of reddened high-$z$ galaxies is clearly
seen in the data. 

\begin{figure}[b!] 
\centerline{\epsfig{file=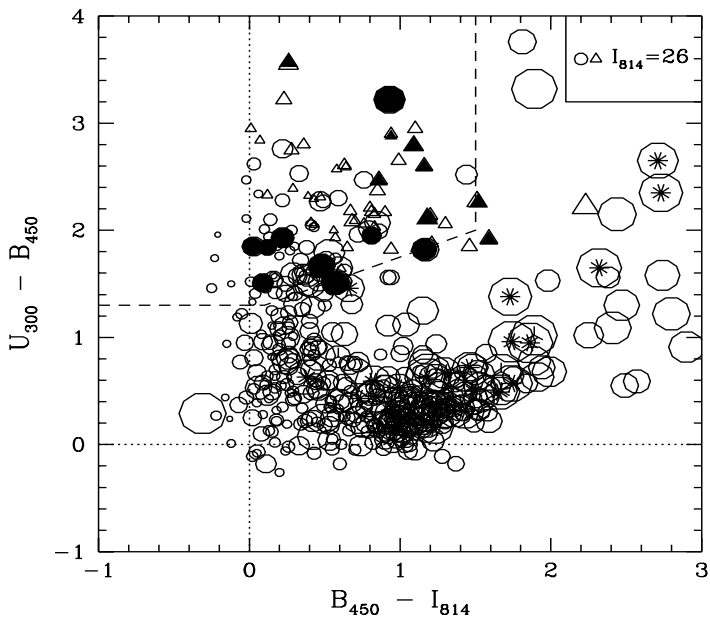,height=4.0in,width=4.5in}}
\vspace{10pt}
\caption{Color-color plot of galaxies in the HDF with $B<27$. Objects
undetected in $U$ (with $S/N<1$) are plotted as triangles at the 1$\sigma$
lower limits to their $U-B$ colors. Symbols size scales with the $I$ mag
of the object. The dashed lines outline the selection region within which we
identify candidate $2<z<3.5$ objects. Galaxies with spectroscopically confirmed
redshifts within this range are marked as solid symbols. Galaxies with
confirmed redshifts lower than $z=2$ are marked as asterisks. 
}
\label{fig2}
\end{figure}

Assuming the redshift interval $2<z<3.5$ has been uniformly probed, 
I have plotted in Figure 1 the SFR function of the $U$ dropouts in the 
HDF. From the observed $V$ magnitudes of our sample, a ``directly-observed'' 
SFR density at $\langle z\rangle=2.75$ of ${\dot \rho_*}=
10^{-1.66}\msunits$ is computed. As only a short segment of the LF can be
determined, however, it is dangerous to fit the usual Schechter function to get
a better estimate of the total density of starbirth. What I have done instead
is to show in Figure 1 that the $z\sim 1$ Schechter function with a
normalization $\phi^*$ {\it lower} by a factor of 1.6 provides a good fit to
the HDF points. The integration of this function over all luminosities yields 
a star formation density about 1.45 larger than the directly-observed one.
In the following, I will adopt the intermediate value 
\[
{\dot \rho_*}=10^{-1.57}\, {\rm M_\odot\,yr^{-1}\,Mpc^{-3}}
\]
as our best determination of the SFR density at $\langle z\rangle=2.75$, 
and assign to it an uncertainty of $\pm 0.15$ in the log.
A comparison between the comoving space density of the HDF $U$ dropouts
brighter than $V=25.5$ and that derived from ground-based statistic
[13] (${R}<25.5$) at redshift $\langle z\rangle=3.25$, shows
good agreement, after accounting for the fact that one probes $\sim 0.3$
mag fainter in the galaxy LF at the HDF average redshift. The combination 
of the ground-based Lyman-break galaxy survey with the $U$ dropout
sample provides a better sketch of the distribution of SFR for galaxies at
$z\approx 3$ [22]. 

\bigskip {\bf Blue Dropouts} \qquad 
In analogous way, the $B_{450}$ passband allows the selection of candidate
star-forming galaxies in the interval $3.5<z<4.5$. We have identified $\sim 20$
$B$ dropouts to $V<28$ in the $\bv$ vs. $\vi$ plane [18]. The
brightest one has been tentatively confirmed to be at $z=4.02$, consistent with
the photometric predictions [23]. From the observed $I$
magnitudes of our sample, a SFR density of 
\[
{\dot \rho_*}=10^{-2.06\pm 0.2}\, {\rm M_\odot\,yr^{-1}\,Mpc^{-3}}
\]
is obtained at $\langle z\rangle=4.0$, well below the $\langle z\rangle=2.75$
value. The error bar reflects the uncertainties present in the volume
normalization and in the color selection given the lack of spectroscopic
confirmations. 

\section*{Cosmic Metal Production}
The (rest-frame) radiation flux below $3000\,$\AA\ is a very good measurement
of the instantaneous ejection rate of heavy elements ($Z\ge 6$), since both are
directly related to the number of massive stars [24]: the same
stars with $m>10\msun$ that manufacture and return most of the metals to the
ISM as Type II supernovae also dominate the UV light. Contrary to the
conversion from UV to SFR, which is a sensitive function of the IMF slope, the
UV-to-metal conversion efficiency is fairly insensitive to the assumed IMF,
since the increased metal yield [25] from high mass stars is
compensated for by a similar increase in the production of UV photons. 

The rate of ejection of newly synthesized material per unit comoving volume as
a function of redshift, $\dot \rho_Z$, is shown in Figure 3, together with a
sketch of the cosmic star formation history in galaxies [18].
When combined with the ground-based data, the HDF results appear
consistent with the existence of a peak in the universal metal production rate
in the redshift range $1<z<2$, in agreement with inferences from quasar 
absorption systems [26]. The plot suggests
that, while the conversion of gas into stars must have been extremely efficient
at intermediate redshifts, and galaxies have largely exhausted their reservoirs
of cold gas at the present-epoch, there must be a mechanism which prevent the
gas within virialized dark matter halos to radiatively cool and turn into stars
at $z \gtrsim3$. 

We may at this stage try to establish a cosmic timetable for the production of
metals in relatively bright galaxies, keeping in mind the inherent
uncertainties associated with the estimates given above. The mass density
of heavy elements observed today in ``normal'' massive galaxies is 
\[ 
\rho_Z(0)=Z_*\rho_B(0)\langle{M\over L_B}\rangle=6\pm 3\times 10^6 
({Z_*\over Z_\odot})\munits
\]
where $\rho_B(0)$ is the local blue light density, and
$\langle M/L_B\rangle\approx 3$ is the mean mass-to-blue
light ratio of visible matter in solar units. Although a baryonic mass several
times larger than the luminous mass may be present in the Galactic halo,
metal-rich halo material would be mixed into and over-enrich the disk. Hence,
if a substantial amount of metals are missing from our census, they are most
likely hidden in the intergalactic gas. 

If we define two characteristic epochs of star and element formation in
galaxies, $z_*$ and $z_Z$, as the redshifts by which half of the current
stellar and metal content of galaxies was formed, then a straightforward
integration of the curve plotted in Figure 3, together with the fact that most
of the stars in the inner luminous parts of galaxies are metal rich, imply
$z_*\lesssim z_Z\approx 1$, or in other words that a significant fraction of the
current metal content of galaxies was formed relatively late by late-type
systems, on a timescale of about 8 Gyr. (Note that, contrary to the measured
number densities of objects and rates of star formation, the total metal mass
density produced is independent of the assumed cosmology.)

\begin{figure}[b!] 
\centerline{\epsfig{file=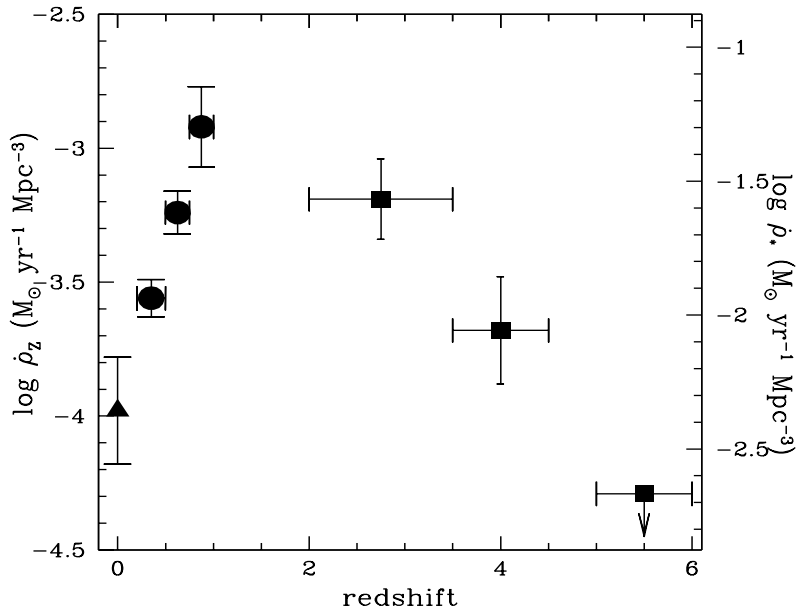,height=4.0in,width=4.5in}}
\vspace{10pt}
\caption{Element and star formation history of the universe. The upper limit
at $\langle z\rangle=5.5$ reflects the paucity of $V$ dropouts in the HDF.}
\label{fig3}
\end{figure}

This suggests the possibility that we may be observing in the redshift range
$z=0-1$ the conversion into stars of gaseous galactic disks. Pure \HI disks may
be assembled at some higher redshift, and disk gas continuosly replenished as a
result of ongoing infall from the surrounding hot halo. From stellar population
studies we also know that about half of the present-day stars -- hence metals
-- are contained into spheroidal systems, i.e., elliptical galaxies and spiral
galaxy bulges, and that these formed early and rapidly (see, however,
[27]), experiencing a bright starburst phase at high-$z$. Where are
these protospheroids? 

The space density of bright ellipticals today is $\phi(>L_*)\approx 4.5\times
10^{-4}\,$Mpc$^{-3}$ [6].  If a significant fraction of their
stellar population formed in a single burst of duration 1 Gyr early in the
history of the universe, a comparable number of galaxies should be observed at
high-$z$ while forming stars at rates in excess of about $100\msun$ yr$^{-1}$.
From the Lyman-break galaxy sample, however, the space density of high
star-forming galaxies at $z\sim 3$ is about 50 times lower [13].
Hence there is a serious deficit of very bright objects relative to the
expectations of the standard early-and-rapidly-forming picture for spheroidal
systems. At the star formation density levels inferred from the HDF images,
about 1/3 of the observed mass density of metals at $z=0$ would have been
formed during the ``spheroid epoch'' at $z\gtrsim 1.5$. 

Finally, it is only fair to point out that all the values derived above
should be considered as lower limits, as newly formed stars which are
completely hidden by dust would not contribute to the H$\alpha$ or UV
luminosity. It is a fact that, at the present-epoch, we are
underestimating the SFR density by about a factor of 2, as a ``typical''
optically-selected spirals emits 30\% of its energy in the FIR region
[28]. On the other hand, since the metals we observe being formed
are a substantial fraction of the entire metal content of massive galaxies, it
appears that -- on average --  star formation regions remain largely unobscured
by dust throughout much of galaxy formation. The opposite can be true only if
galaxies eject a large amount of heavy elements in the intergalactic medium.

\end{document}